# Improved normal-boundary intersection algorithm: a method for energy optimization strategy in smart buildings

Jia Cui [a], Jiang Pan [a], Shunjiang Wang [b], Martin Onyeka Okoye [a], Junyou Yang [a], Yang Li [c], Hao Wang [a]

a School of Electrical Engineering, Shenyang University of Technology, Shenyang 110870, Liaoning Province, China; cuijiayouxiang@163.com(C.J.); jpan801@163.com(J.P.); junyouyang@sut.edu.cn(J.Y.); martinokoye@yahoo.com; wanghao2000wh@163.com(H.W.)

b State Grid Liao Ning Electric Power Company, Shenyang 110015, Liaoning Province, China; ( Email: wangshunjiang@163.com )

c School of Electrical Engineering, Northeast Electric Power University, Jilin 132012, China (e-mail:liyang@neepu.edu.cn)

**Abstract**

With the widespread use of distributed energy sources, the advantages of smart buildings over traditional buildings are becoming increasingly obvious. Subsequently, its energy optimal scheduling and multi-objective optimization have become more and more complex and need to be solved urgently. This paper presents a novel method to optimize energy utilization in smart buildings. Firstly, multiple transfer-retention ratio (TRR) parameters are added to the evaluation of distributed renewable energy. Secondly, the normal-boundary intersection (NBI) algorithm is improved by the adaptive weight sum, the adjust uniform axes method, and Mahalanobis distance to form the improved normal-boundary intersection (INBI) algorithm. The multi-objective optimization problem in smart buildings is solved by the parameter TRR and INBI algorithm to improve the regulation efficiency. In response to the needs of decision-makers with evaluation indicators, the average deviation is reduced by 60% compared with the previous case. Numerical examples show that the proposed method is superior to the existing technologies in terms of three optimization objectives. The objectives include 8.2% reduction in equipment costs, 7.6% reduction in power supply costs, and 1.6% improvement in occupants' comfort.

## Nomenclature

| | | | |
|---|---|---|---|
| $C_l$ | Total cost of the photovoltaic power generation (¥) | $P_{Gf}$ | Photovoltaic power (MW) |
| $C_w$ | Total cost of the wind power generation (¥) | $P_{Gj}$ | Removable load (MW) |
| $C^{grid}$ | Unit cost of power supply cost (¥) | $P_{Jh}$ | The grid (MW) |
| $ceil$ | The rounding up function | $P_{Kd}$ | Schedulable load(MW) |

| Symbol | Description | Symbol | Description |
|---|---|---|---|
| $d_i^+$ | Mahalanobis distances between the $i$ th non-inferior solution and the positive ideal points | $P_{Kq}$ | Critical load (MW) |
| $d_i^-$ | Mahalanobis distances between the $i$ th non-inferior solution and negative ideal points | $P_k^l$ | Pareto frontier on one of the coordinate axis planes |
| $d_{mn}^c$ | Length of the $c$ th dividing line segment | $P_{u.f}$ | The building's usage of wind power generation (MW) |
| $d_{set}$ | Minimum distance between two points on the Pareto frontier in the original normal-boundary intersection (NBI) | $P_{u.g}$ | The building's usage of photovoltaic power generation (MW) |
| $F_a, F_b$ | Normalized functions in one of the planes | $P_{wm}$ | Monthly average generated power (MW) |
| $\overline{F_a}, \overline{F_b}$ | Normalization of objective functions | $P_{s.f}$ | Total wind power generation of the buildings (MW) |
| $f_1(i)$ | One of the objective functions | $P_{s.g}$ | Total photovoltaic power generation of the buildings (MW) |
| $f_2(i)$ | One of the objective functions | $S_D$ | Occupants' comfort (%) |
| $f_3(i)$ | One of the objective functions | $Th_i$ | The relative closeness of non-inferior solution |
| $G_0(i)$ | Inequality constraints | $T_{in}$ | Indoor temperature (°C) |
| $G(P_k^l)$ | The second coordinates obtained after satisfying its inequality constraints in the plane | $u^+$ | The positive ideal points |
| $H_0(i)$ | Equality constraints | $u^-$ | The negative ideal points |
| $H(P_k^l)$ | The second coordinates obtained after satisfying its equality constraints in the plane | $w_1$ | Wind power operating cost factor |
| $i$ | Normalized quantity | $w_2$ | Maintenance cost factor |

| Symbol | Description | Symbol | Description |
|---|---|---|---|
| $K_\alpha, K_\beta$ | The action coefficient matrix of electrical connection | $x, y, z$ | Coordinates of the uniform point of the projection surface |
| $l$ | Photovoltaic power generation factor | $\alpha, \beta$ | The transfer-retention ratio (TRR) |
| $m$ | Number of special buildings | $\omega$ | Weight of the diagonal matrix |
| $n$ | Number of ordinary buildings | $\omega_{mn}^c$ | Weight interval |
| $P_0$ | Instantaneous value of the wind power recorded every 15 minutes (MW) | $\overline{\omega}_{mn}$ | Average of the weights of the three coordinate axis planes |
| $P_1$ | Total power of each special building (MW) | $\omega_{mn}^{\min}$ | Expansion factor |
| $P_2$ | Total power of each ordinary building (MW) | $\rho$ | The point on the plane mapped to the original Pareto |
| $P_a^m$ | The abscissa corresponding to point $P_a$ | $\rho_b$ | Coordination parameter of the load uncertainty and electrical connection in the building |
| $P_b^n$ | The ordinate corresponding to point $P_b$ | $\gamma_l$ | Combined effect coefficient of photovoltaic power and wind power in the building |
| $P_c$ | Maximum output power of photovoltaic cells (MW) | $\lambda_{ab}$ | Weight of one of the planes |
| $P_{Fl}$ | Wind power (MW) | $\Sigma^{-1}$ | The inverse of the objective covariance matrix |

## 1. Introduction

In recent years, with the widespread use of distributed energy, the advantages of smart buildings over traditional buildings have become more obvious [1]. Also, their optimized scheduling and energy management have widely received attention [2]. The cooperative interaction within buildings formed by multiple smart buildings can achieve efficient use of renewable energy [3]. The smart building is defined as a new type of building equipped with distributed renewable energy generation facilities [4]. The smart building is proposed as the relationship between the power supply, the grid, load, and energy storage where the energy storage is more closely related [5]. The difficulty in the energy optimization and control of smart building group is the regulation of distributed renewable energy [6].

For the smart building with distributed renewable energy, literature [7-9] adopted the model prediction method to model its load in order to improve the real-time controllability of its integrated energy. Based on the smart building load model prediction, literature [10] further forecasts renewable energy generation. Literature [11，12] analyzed the distributed clean energy regulation method based on demand response and established the multi-objective optimization model of the smart building. In literature [13], The multi micro-grid system based on cogeneration is analyzed. Literature [14] analyzes the micro-grid system based on distributed energy, energy storage system and demand response. However, the effects of wind and solar correlation, load uncertainty, and electrical connection on smart building control were not fully considered. Furthermore, the optimization and regulation of complex distributed energy need to be improved. Therefore, further exploration is needed in this aspect.

In the aspect of finding solutions to multi-objective optimization problems, literature [15,16] used particle swarm optimization algorithm to solve multi-objective optimization problems. However, it is difficult in solving complex multi-objective optimization problems. The distributed multi-objective optimization algorithm was proposed in literature [17]. The multi-objective optimization algorithm based on the weighted sum method was proposed in literature [18]. The multi-objective optimization algorithm based on Pareto frontier is adopted in literature [19]. This is improved in literature [20,21] to meet the requirements of multiple working conditions. In literature [22], the adjust uniform axes method (AUAM) algorithm is used to improve the normal-boundary intersection (NBI) algorithm. This makes Pareto frontier widely distributed and uniform. Literature [23] combines multi-modal multi-objective evolutionary algorithm with refined coding genetic algorithm to find the optimal schedule. Literature [24] combines multi-objective optimization algorithm with distributed consensus mechanism. However, the NBI algorithm and AUAM are difficult to get a widely distributed and uniform Pareto frontier when dealing with complex multi-dimensional problems. Also, they are not close enough to the real Pareto frontier. Therefore, a more extensive and uniform Pareto frontier algorithm is needed to solve multi-objective optimization problems.

Firstly, this paper presents a model of buildings' internal energy consumption in conjunction with wind power and photovoltaic power generation. The total load model has been set. Also, a series of constraints and the cost price of each unit have been established. Secondly, for the analysis of smart buildings using the INBI algorithm improved by the adaptive weight sum (AWS) method, the AUAM and Mahalanobis distance double base point method are used as multi-objective optimization algorithms. This is to optimize equipment costs, power supply costs, and occupants' comfort. Finally, under different cases, the optimal results in each case are calculated according to the above algorithm.

The contributions of this paper are summarized as follows:

(1) For the first time, the evaluation index of building's regulation named transfer-retention ratio (TRR), including wind, solar, and air conditioning-combined dispatching system, is proposed. The TRR considers the combined effect of wind power and photovoltaic electrical connection and load uncertainty, making it interpretively better for distributed energy systems.

(2) An optimization algorithm based on the NBI algorithm is proposed. This is based on the improved AWS algorithm to increase the feasible solution across latitude. It utilizes the AUAM algorithm to solve the problem of wide and uniform distribution of the Pareto frontier. It also utilizes the Mahalanobis distance double base point method as the screening criteria to further make the Pareto frontier widely and evenly distributed.

The rest of the paper is organized as follows: Section 2 introduces the smart building system with wind, solar, and air conditioning system. It proposes the building's adjustment evaluation indicators and presents the corresponding optimization calculations. Section 3 improves the NBI method to provide theoretical support for the case analysis in Section 4. Finally, a conclusion is given in Section 5.

## 2. A control index considering comprehensive coordination of wind power, solar power, and air conditioning system in the building

*2.1 The smart building model*

The smart building contains many units that can be divided into two parts in terms of energy exchange [25]. As shown in Figure 1, the grid, wind power generation, and photovoltaic power generation are the power supplies, and the various types of loads represent power consumption.

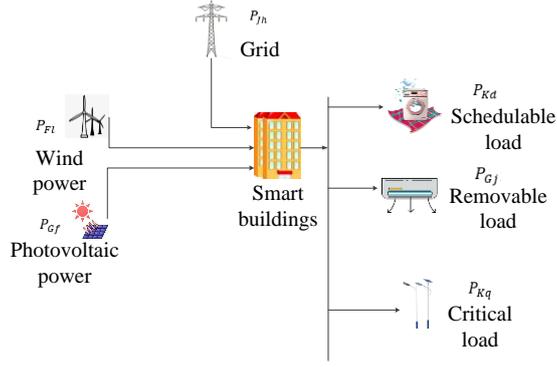

**Fig. 1.** Various types of units in smart buildings

Figure 1 shows the relationship between the various units in the building. The main goal of this paper is to achieve the optimization of equipment cost, power supply cost, and occupants' comfort [26]. Equipment cost is mainly composed of wind power and photovoltaic power generation including equipment installation and maintenance, operation loss compensation, etc.

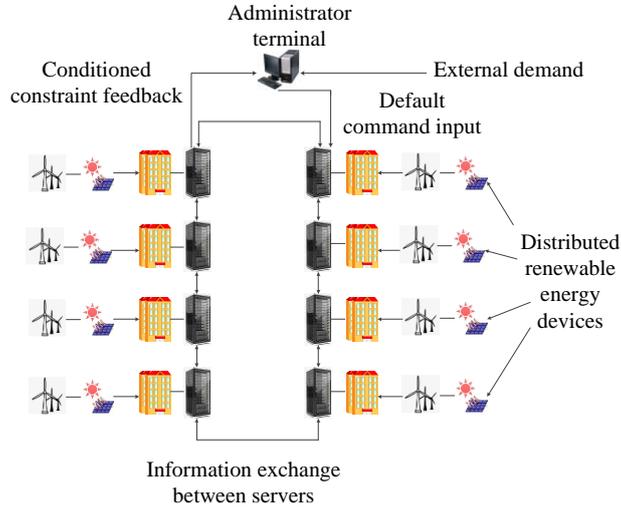

**Fig. 2.** Distributed energy in smart buildings

As shown in Figure 2, in the smart building construction, each building is equipped with independent wind power generation and photovoltaic power generation facilities and independent energy storage equipment [27]. The energy storage equipment between each building is interconnected.

The meanings of the three quantities regulated in this paper are as follows: equipment cost is the total cost of the establishment and maintenance of all equipment in a month; power supply cost is the electricity cost of all users' loads in the building in a month; occupants' comfort is the average of all the occupants' comfort values in a month.

*2.2 Various component models and constraints in the smart building*

It is necessary to consider the cost, loss, and generated electrical energy to reduce the amount of power supplied by the grid [28,29]. Thus, a wind power cost model is established as:

$$C_w + C_l = w_1 P_0 + w_2 P_{wm} + l P_c \qquad (1)$$

where $C_w$ is the total cost of the wind power generation, $C_l$ is the total cost of the photovoltaic power generation, $P_0$ is the instantaneous value of the wind power recorded every 15 minutes, $P_{wm}$ is the monthly average power generated, $P_c$ is the maximum output power of photovoltaic cells, $w_1$ and $w_2$ are the wind power operating cost factor and maintenance cost factor, respectively, and $l$ is the cost factor of photovoltaic power generation and maintenance. In this paper, referring to the calculation standard of the renewable energy cost in buildings and considering the characteristics of renewable energy storage, the formula in the relevant documents is simplified and presented.

The unit cost of the power supply, $C^{grid}$, is given by (2) [30].

$$C^{grid} = \begin{cases} 0.3, t = t_1 \\ 0.7, t = t_2 \end{cases} \quad (2)$$

where $t_1$ is the valley time which is from 23:00 hours to 06:00 hours of the following day. $t_2$ is the peak time which is from 06:00 hours to 23:00 hours daily. This is the scientific standard of electricity charge for general buildings in North China.

The occupants' comfort, $S_D$, is given by (3) [31].

$$S_D = [1 - |T_{in} - 26| \times 0.25] \times 100\%, \ T_{in} \in [22, 30] \quad (3)$$

where, $T_{in}$ is the indoor temperature. In this paper, thermal comfort is selected from the definition formula of total comfort, and the complex formula is simplified to facilitate the calculation and understanding.

The power balance constraints are given by (4) [32]:

$$\begin{cases} P_{Jh} + P_{Fl} + P_{Gf} = \sum_{i=1}^{z} P_{Kd}(k) + P_{Gj} + \sum_{j=1}^{Z} P_{Kq}(k) \\ \underline{P_{Fl}} \leq P_{Fl} \leq \overline{P_{Fl}}, \underline{P_{Gf}} \leq P_{Gf} \leq \overline{P_{Gf}} \end{cases} \quad (4)$$

where $P_{Jh}$ is the power exchanged with the grid, $P_{Fl}$ is the wind power output power, $P_{Gf}$ is the photovoltaic-generated power, $P_{Kd}$ is the schedulable load power, $P_{Gj}$ is the critical load, and $P_{Kq}$ is the switchable load power.

*2.3 The modeling of the control quantity based on wind, solar, and air conditioning dispatch system in the smart building*

The smart building adopts the distributed renewable energy architecture, which can improve the effective utilization rate of renewable energy. Some of its renewable energy can be directly connected with users, reducing the power consumption brought by electric energy storage. Compared with the traditional mode of using the total energy storage battery, it has advantages [33]. In large-scale dispatching, in order to make the users in the building obtain more stable and effective power, it is necessary to consider that the renewable energy resources in the region should be evenly distributed to each building as far as possible. which This can not only improve the effective utilization of power but also take into account the satisfaction of users.

In multiple smart buildings, each building has its own distributed renewable energy equipment. To optimize its regulation, the corresponding regulation index requires to be described [34]. The variables, $\alpha$ and $\beta$, are defined as the total photovoltaic power generation ratio and the total wind power generation ratio of the smart building, respectively. In this paper, these two variables are defined as the ratio of renewable energy output and reserve of a smart building named TRR. Their values are recorded every 15 minutes, thus 96 values in a day. They are as expressed in (5) and (6), respectively.

$$\alpha = \rho_b \frac{P_{u.g} + \gamma_l (P_{s.f} - P_{u.f})}{P_{s.g}} \times \left( -\ln \frac{P_{s.g} \times P_{u.g}}{\sum P_{s.g}^2} \right) \quad (5)$$

$$\beta = \rho_b \frac{P_{u.f} + \gamma_l (P_{s.g} - P_{u.g})}{P_{s.f}} \times \left( -\ln \frac{P_{s.f} \times P_{u.f}}{\sum P_{s.f}^2} \right) \quad (6)$$

where $P_{u.g}$ and $P_{u.f}$ are the building's usage of photovoltaic power generation and wind power generation, respectively; $P_{s.g}$ and $P_{s.f}$ are the total photovoltaic power generation and wind power generation of the buildings, respectively whose values are recorded every 15 minutes, thus 96 values in a day; $\gamma_l$ is the combined effect coefficient of photovoltaic power and wind power in the building; $\rho_b$ is the coordination parameter of the load uncertainty and electrical connection in the building.

The wind power equation is given by (7).

$$P_{s.f} = \int_{v_i}^{v_{i+1}} P_{Fl}(v) dv \quad (7)$$

The photovoltaic equation power is given by (8).

$$P_{s.g} = \int_{t_u}^{t_{u+1}} P_{Gf}(t) dt \quad (8)$$

After defining the parameter TRR, to solve the multi-objective optimization problem considering the effect of electrical connection, this paper integrates TRR constraint into the INBI algorithm. The formula of the parameter, TRR, is given as:

$$\begin{cases} \min \sum K_\alpha |\alpha_i - \alpha_j| \\ \min \sum K_\beta |\beta_i - \beta_j| \end{cases} \quad (9)$$

where $K_\alpha$ and $K_\beta$ are the action coefficient matrices of electrical connection, $\alpha_i$ and $\alpha_j$ are the values of two observation points in the parameter, TRR, of photovoltaic, $\beta_i$ and $\beta_j$ are the values of two observation points in the parameter, TRR, of wind power.

The energy regulation of smart buildings is complex. The regulation ability of distributed renewable energy is improved by the parameter TRR. Moreover, in the face of the optimization problem, the boundary conditions of the multi-objective optimization algorithm can be added to the parameter to further improve the ability to solve the optimization problem.

## 3. Improved INBI algorithm-based solution approach

Based on the original NBI algorithm [35], this paper improves its optimization degree by the integration of the following algorithms: AWS, AUAM, and Mahalanobis distance double base point method. The algorithm sequential steps are as follows:

(a) The original NBI algorithm optimization model is given;

(b) The modified Pareto frontier is obtained by the AWS algorithm;

(c) The modified Pareto frontier is projected to the three-target Pareto frontier;

(d) The three-target Pareto frontier points are corrected and the non-inferior solutions are selected by AUAM;

(e) The Mahalanobis distance double base point method is used to replace the distance method in the original NBI algorithm to obtain a compromise solution.

The computational and time complexities are showed in figure 3 and 4. The flowchart and advances of INBI algorithm are shown in Figure 3. Based on the improvement of AWS, feasible solutions are added across dimensions to ensure the difference of feasible solutions and avoid homogenization. The wide and uniform problem of Pareto frontier are optimized based on AUAM to achieve the purpose of optimizing screening criteria. Moreover, AWS-AUAM is combined closely. Taking three-dimensional as an example, AWS produces three different step sizes. AUAM needs to be improved to solve the problem of step size adaptation. Multi step size is preferred - minimum, so as to ensure no leakage points and improved detection sensitivity. Considering the influencing factors of personal preference, Markov distance is used to replace Euclidean distance. The traditional evaluation matrix and distance do not consider the correlation. And the positive and negative ideal points are modified based on the weight diagonal matrix to avoid overlapping information affecting the optimal decision results. Mahalanobis distance only considers the evaluation criteria, which is more simple, practical, comprehensive and accurate than other complex distance functions.

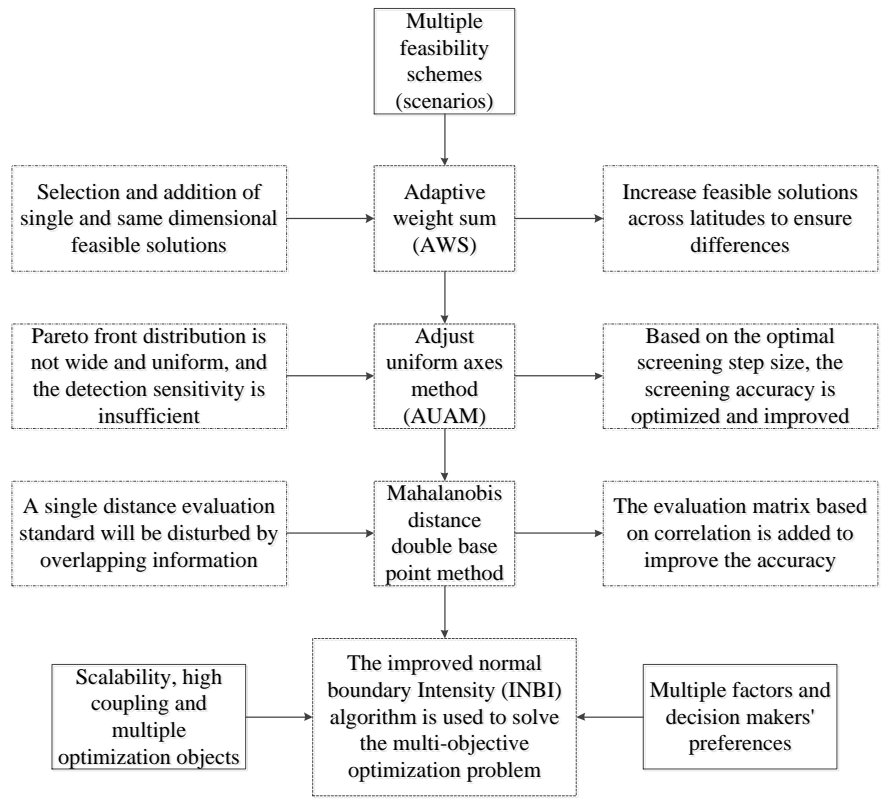

**Fig. 3**. The flowchart and advances of INBI algorithm

A mapping table 1 and map limitation is made to address the limitations, solutions and methods of validation.

**Table 1.** The mapping relationship in this paper

|  | Restrict | Solution | Verification method |
|---|---|---|---|
| The TRR | The effects of wind and solar correlation, load uncertainty, and electrical connection on smart building control were not fully considered. Furthermore, the optimization and regulation of complex | The TRR considers the combined effect of wind power and photovoltaic electrical connection and load uncertainty. Moreover, TRR-INBI can be used to solve the the optimization and regulation of complex | In contrast, the average deviation is reduced by 60%. The result refers to figure 11. |

| | distributed energy need to be improved. | distributed energy | |
|---|---|---|---|
| The INBI algorithm | The NBI algorithm and AUAM are difficult to get a widely distributed and uniform Pareto frontier when dealing with complex multi-dimensional problems. | The INBI algorithm makes the Pareto frontier more extensive and uniform. | For equipment cost, power supply cost, and occupants' comfort, the proposed algorithm is optimized by 8.2%, 7.6%, and 1.6%, respectively. The results refer to table 3-7. |

The flow chart is shown in Figure 4. In this paper, AWS algorithm is used to expand the number of feasible solutions. After the original AWS modifies the two-dimensional Pareto frontier, each point generated in the correction process is retained considering the marginal effect. The original AWS algorithm generates a correction point from a initial point. And each point in the correction process is expanded to B times of the original. As a result, its spatial complexity is changed from o (A) to o (A + BA), and the time complexity is still o (A). The number of points generated by the original NBI algorithm is set as C. When each point is calculated by AUAM algorithm, its spatial complexity and time complexity change from O (C) to o (C + A + BA).

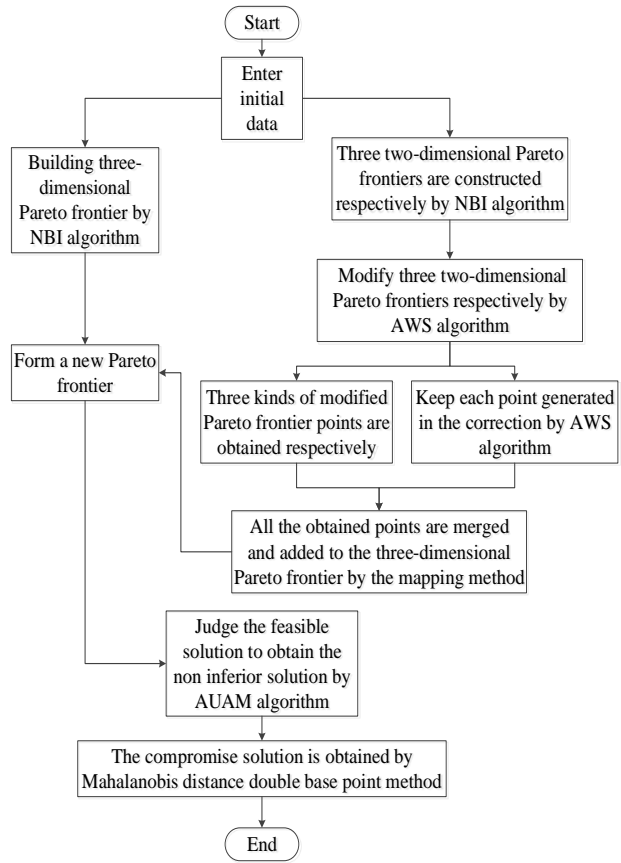

**Fig. 4**. The flow chart of INBI algorithm

*3.1 The original NBI three-objective optimization model*

The original NBI three-objective optimization model is given by (10).

$$\begin{cases} \min_{i \in M} F_0(i) = \{f_1(i), f_2(i), f_3(i)\} \\ M = \{i : H_0(i) = 0, \underline{G_0} \leq G_0(i) \leq \overline{G_0}\} \end{cases} \quad (10)$$

where $f_1(i)$ is the equipment cost; $f_2(i)$ is the power supply cost; $f_3(i)$ is the occupants' comfort; $H_0(i)$ and $G_0(i)$ are the equality constraints and inequality constraints, respectively; $i$ is the normalized quantity. Similar to this is the dual objective NBI optimization model.

*3.2 The modified Pareto frontier obtained by the AWS algorithm.*

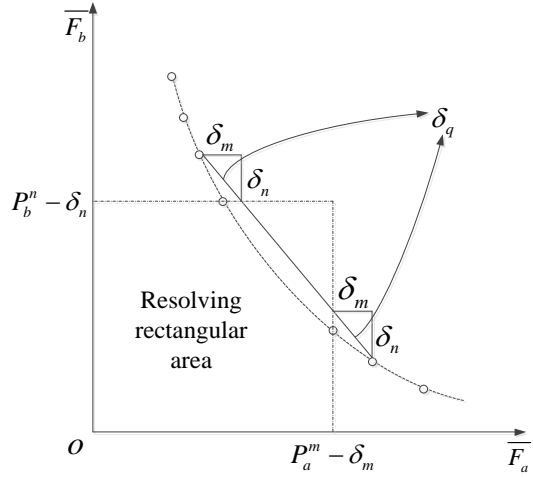

**Fig. 5.** Modified Pareto frontier obtained by the AWS algorithm in one of the coordinate planes

As shown in Figure 5, $d_{set}$ is set as the minimum distance between two points on the Pareto frontier in the original NBI. In resolving the rectangular area, the neighborhood intersection of the original Pareto point in the range of $2d_{set}$ forms the Pareto frontier point group to be corrected. $\overline{F_a}$ and $\overline{F_b}$ are the normalization of objective functions, $F_a$ and $F_b$, respectively [36].

$$\begin{cases} \min \lambda_{ab} \dfrac{F_a(v) - F_{a\min}}{F_{a\max} - F_{a\min}} + (1 - \lambda_{ab}) \dfrac{F_b(v) - F_{b\min}}{F_{b\max} - F_{b\min}} \\ s.t. \underline{G} \leq G(v) \leq \overline{G}, H(v) = 0 \\ F_a, F_b, \lambda_{ab} \in \{F_1, F_2, \lambda_{12}; F_3, F_4, \lambda_{34}; F_5, F_6, \lambda_{56}\} \\ G \in \{g_1, g_2, g_3\}, H \in \{h_1, h_2, h_3\} \end{cases} \quad (11)$$

$$\begin{cases} \dfrac{F_a(v) - F_{a\min}}{F_{a\max} - F_{a\min}} \leq P_a^m - \delta_m \\ \dfrac{F_b(v) - F_{b\min}}{F_{b\max} - F_{b\min}} \leq P_b^n - \delta_n \\ \lambda_{ab} \in \left\{0, \dfrac{1}{\omega_{mn}}, \dfrac{2}{\omega_{mn}}, ..., \dfrac{\omega_{mn} - 1}{\omega_{mn}}, 1\right\} \\ \omega_{mn}^c = ceil\left(\dfrac{d_{mn}^c}{\delta_q}\right) - 1, \delta_q = \sqrt{\delta_m^2 + \delta_n^2} \end{cases} \quad (12)$$

where, $F_a$ and $F_b$ are the normalized functions in one of the planes, respectively; $F_{a\min}$ and $F_{b\max}$ are the minimum and maximum values of $F_a$ and $F_b$, respectively after the $F_{a\min}$ is separately calculated; $F_{a\max}$ and $F_{b\min}$ are the maximum and minimum values of $F_a$ and $F_b$, respectively after the $F_{b\min}$ is separately calculated; $\lambda_{ab}$ is the weight of one of the planes; $\omega_{mn}^c$ is the weight interval; $P_a^m$ and $P_b^n$ are the abscissa corresponding to point $P_a$ and the ordinate corresponding to point $P_b$, respectively; $d_{mn}^c$ is the length of the $c$th dividing the line segment; $P_a P_b$, $\delta_q$, $\delta_m$, and $\delta_n$ are the feasible regions of the distance; $\delta_q$ is $2d_{set}$; ceil is the rounding up function; $F_1$, $F_2$, $\lambda_{12}$, $g_1$, and $h_1$ are the parameters and constraints in the $xoy$ plane, respectively; $F_3$, $F_4$, $\lambda_{34}$, $g_2$, and $h_2$ are the parameters and constraints in the $yoz$ plane, respectively; $F_5$, $F_6$, $\lambda_{56}$, $g_3$, and $h_3$ are the parameters and constraints in the $zox$ plane, respectively.

Based on the improved AWS, feasible solutions are added across dimensions to ensure a difference with feasible solutions and avoid homogenization.

*3.3 The modified Pareto frontier projected to the three-target Pareto frontier*

Each point projected to the front of the three-objective Pareto must meet the constraints of the original three-objective NBI model as:

$$\begin{cases} \underline{G_0} \leq G_0(P_k^l, G(P_k^l), \rho) \leq \overline{G_0} \\ H_0(P_k^l, H(P_k^l), \rho) = 0 \end{cases} \quad (13)$$

where $P_k^l$ is the first coordinate of the new Pareto frontier on one of the coordinate axis planes; $G(P_k^l)$ and $H(P_k^l)$ are the second coordinates obtained after satisfying its inequality constraints and equality constraints in the plane, respectively; $\rho$ is the point on the plane mapped to the original Pareto. The third coordinate on the leading edge satisfies the original inequality constraints and equality constraints. $P_k^l$, $G(P_k^l)$, and $\rho$ satisfy the normalization in the original NBI algorithm, and the modified Pareto frontier is obtained from this.

*3.4 The corrected three-target Pareto frontier and the non-inferior solution selected by AUAM*

To ensure that the Pareto frontier is widely and evenly distributed in the three-dimensional space, the modified Pareto frontier is selected by the AUAM [22] in this paper.

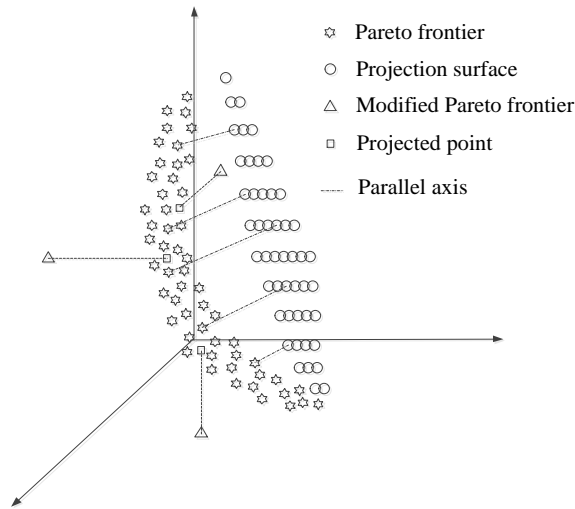

**Fig. 6.** Selection of modified three-target Pareto frontier by AUAM

As shown in figure 6, the projected point is obtained by projecting the modified Pareto frontier points on the three coordinate axis planes. The modified Pareto frontier obtained by the AWS algorithm in the three surfaces is projected into the original Pareto frontier to form the modified three-target Pareto frontier. Then, the parallel axes are used between the Pareto frontier and the projection surface as the selection criteria.

To select the Pareto frontier in the original NBI algorithm model and the three-objective modified Pareto frontier uniformly, the two Pareto frontiers are combined. The three optimization goals are achieved in the optimization algorithm separately to obtain the target vector of the three extreme solutions and the maximum value of each target. The target value is normalized separately as given by (14).

$$s_j = \frac{f_j - f_{j\min}}{f_{j\max} - f_{j\min}}, j=1,2,3 \quad (14)$$

where $f_j$ is the original value of the $j$ th target, $s_j$ is the normalized value of the $j$ th target, $f_{j\max}$ and $f_{j\min}$ are the maximum and minimum values of the $j$ th target, respectively.

It is stipulated that $S_A$, $S_B$, and $S_C$ are the target vectors of the three extremum solutions respectively. The plane determined by the points $S_A$, $S_B$, and $S_C$ is called the insinuation surface, $T_R$, of the Pareto frontier. Points on the projection surface satisfy the constraints, thus:

$$\begin{cases} 0 - \omega_{mn}^{\min} \leq x, y, z \leq 1 + \omega_{mn}^{\min} \\ Ax + By + Cz = 1 \\ \omega_{mn}^{\min} = \min \frac{1}{\omega_{mn}} \end{cases} \quad (15)$$

where $x, y, z$ are the coordinates of the uniform point of the projection surface, $Ax+By+Cz=1$ is the projection surface equation, and $\omega_{mn}^{\min}$ is the expansion factor. In this paper, we take the minimum weight interval.

After normalization, the axes are defined, and the axes for the modified three-target Pareto frontier should be necessarily given to ensure that all Pareto frontier points are taken into account. The axis parallel vector is given by (16).

$$\vec{e} = [e_1, e_2, e_3], e_1, e_2, e_3 > 0 \quad (16)$$

The mathematical model of the intersection point between the parallel axis and the Pareto frontier is given by (17).

$$\begin{cases} \max u_\theta \\ s.t. \begin{cases} \bar{s}_{\theta 1} - \hat{s}_{\theta 1} = e_1 u_\theta \\ \bar{s}_{\theta 2} - \hat{s}_{\theta 2} = e_2 u_\theta \\ \bar{s}_{\theta 3} - \hat{s}_{\theta 3} = e_3 u_\theta \\ I(\hat{s}_{\theta 1}, \hat{s}_{\theta 2}, \hat{s}_{\theta 3}) \leq 0, \theta \in N \end{cases} \end{cases} \quad (17)$$

where $\bar{s}_\theta = [\bar{s}_{\theta 1}, \bar{s}_{\theta 2}, \bar{s}_{\theta 3}]$ is the uniform point of the projection surface corresponding to the axis, $D_\theta$, $\hat{s}_\theta = [\hat{s}_{\theta 1}, \hat{s}_{\theta 2}, \hat{s}_{\theta 3}]$ is the intersection point of the axis, $D_\theta$, and the Pareto frontier, $I$ is the model constraint, and $N$ is the axis set.

After defining the parallel axis, the selection of the non-inferior solutions in the NBI algorithm is further achieved with higher accuracy. $s_\theta$ is projected along with the vector **e** to the surface, $T_R$, to obtain $\tilde{s}_\theta$. The Euclidean distances are calculated for $\bar{s}_\theta$ and $\tilde{s}_\theta$, and the shortest distance point is defined as $\tilde{s}_\varepsilon$. The range of the domain that meets $\bar{s}_\theta$ is maintained, thus:

$$\begin{cases} \left|\bar{s}_{\theta,p} - \tilde{s}_{\varepsilon,p}\right| \leq \Delta d_p, \theta, \varepsilon \in N; p = 1, 2, 3 \\ \Delta d_1, \Delta d_2 = \dfrac{1 + 2\omega_{mn}^{\min}}{\bar{\omega}_{mn}} \\ \Delta d_3 = \max\left[\dfrac{(1 + 2\omega_{mn}^{\min})A}{\bar{\omega}_{mn} \cdot C}, \dfrac{(1 + 2\omega_{mn}^{\min})B}{\bar{\omega}_{mn} \cdot C}\right] \\ \bar{\omega}_{mn} = \dfrac{\sum \omega_{mn}}{3} \end{cases} \quad (18)$$

where $\bar{\omega}_{mn}$ is the average of the weights of the three coordinate axis planes.

Based on AUAM, the problem of a wide and uniform Pareto frontier is solved to optimize the screening accuracy. In addition, AWS-AUAM is tightly integrated. Taking 3D as an example, AWS would produce three different step sizes. It is necessary to improve AUAM to solve the problem of step size adaptation. Multi-step size selection is the best and minimum to ensure no missing points and improve the detection sensitivity.

*3.5 The compromise solution obtained by the Mahalanobis distance double base point method*

In the original NBI algorithm, the Euclidean distance double base point method is used as the Pareto frontier selection criterion. Based on the Euclidean distance, the correlation between the optimization objectives is not considered [37]. In this paper, the Mahalanobis distance double base point method is used instead of the original method. This is to avoid the influence of the optimization target leading to repeated information in the results. The specific steps are as follows:

1) Establishment of an evaluation matrix.

The element $u_{ij}$ of the evaluation matrix $U$ is the normalized value of the $j$ th target of the $i$ th non-inferior solution. The normalization method is shown in the formula (14).

2) Determination of the positive and negative ideal points as shown in (19).

$$\begin{cases} u^+ = [u_1^+, u_2^+, u_3^+] \\ u^- = [u_1^-, u_2^-, u_3^-] \end{cases} \quad (19)$$

where $u^+$ are the positive ideal points, $u^-$ are the negative ideal points. Each element of $u^+$ corresponds to the minimum value of the $j$ th target normalization value, and each element of $u^-$ corresponds to the maximum value of the $j$ th target normalization value.

3) Calculation of the Mahalanobis distance between the non-inferior solutions and the two kinds of ideal points is given by (20).

$$\begin{cases} d_i^+ = \sqrt{(u_i - u^+)^T \omega^T \Sigma^{-1} \omega (u_i - u^+)} \\ d_i^- = \sqrt{(u_i - u^-)^T \omega^T \Sigma^{-1} \omega (u_i - u^-)} \end{cases} \quad (20)$$

where, $d_i^+$ and $d_i^-$ are the Mahalanobis distances between the $i$ th non-inferior solution and the positive and negative ideal points, respectively; $u_i$ is the target normalized value vector of the $i$ th non-inferior solution; $\omega = diag(\omega_1, \omega_2, \omega_3)$ is the weight diagonal matrix; $\omega_j$ is the weight of the $j$ th objective representing the decision maker's subjective preference information; $\Sigma^{-1}$ is the inverse of the objective covariance matrix.

4) Calculation of the relative closeness, $Th_i$, of all non-inferior solutions is given by (21).

$$Th_i = \frac{d_i^+}{d_i^+ + d_i^-} \tag{21}$$

Each $Th_i$ represents each of the non-inferior solutions. The smaller the value of $Th_i$, the better the non-inferior solution.

Considering the influence factors of personal preference, Mahalanobis distance is used instead of Euclidean distance. The traditional evaluation matrix and distance do not consider the correlation. Also, the positive and negative ideal points are corrected based on the weight diagonal matrix to avoid overlapping information which affects the optimal decision results. Mahalanobis distance only considers the evaluation criteria. Compared with other complex distance functions, Mahalanobis distance is more realistic, practical, comprehensive, and accurate.

## 4. Analysis of samples

*4.1 Initial data and cases setting*

In this section, a case study is performed using INBI based on the power grid load data of a region at Liaoning province, China, in August 2018, to determine its effectiveness. The total number of buildings that participated in the calculation is 20, of which 10 are ordinary buildings and 10 are special buildings. The cost factor for wind power generation operation is 2070(¥/MW). The cost factor for wind power maintenance is 0.096(¥/kW). And the cost factor for photovoltaic power generation is 5000(¥/MW). Taking 15 minutes as an observation time interval, a total of 96 observations were made throughout the day, and the average value is taken in every 4 observation points. This is summarized as 24-hour observation data. In this paper, as one of the optimization objectives of the optimization algorithm, the equipment cost includes the assembly cost, operation cost, and maintenance cost of renewable energy equipment. The hardware specifications used in this simulation are as follows: the CPU is AMD Ryzan 5900HX and the memory is DDR4 3200Mhz 32g. The total energy consumption of each building is basically the same. The decision-makers hope that each building has the same amount of renewable energy as much as possible. The other parameters are consistent with the building standard in North China.The data set used in this paper is the native data of a region, because the privacy policy and confidentiality agreement are inconvenient to be disclosed. Papers using this data set include in reference [38].

This algorithm is suitable for the case which includes three or more optimization objectives, and has a clear and independent algebraic relationship between each optimization objective. Firstly, there are existing researches of the two optimization objectives for three-objective or above optimization, which are based on AWS Two-dimensional method, but they are not combined with AUAM. Secondly, the application conditions of AWS require that there should be a functional relationship between two optimization targets. But if the relational function is complex, it is difficult to split. For example, if the implicit function is used, mathematical means such as denominator 1 or numerator 0 should be adopted to achieve the requirements. The best case is A= F (B), not A= F (B,C).

$\lambda_{ab}$ is the weight of one of the planes in AWS algorithm, $\omega_{mn}^{\min}$ is the expansion factor and $\overline{\omega}_{mn}$ is the average of the weights of the three coordinate axis planes in AUAM algorithm.These parameters are determined by the super parameter $\frac{1}{\omega_m}$ representing the step size. The step size is determined by the functional relationship between two optimization objectives on different planes.

As AWS algorithm is sensitive in the selection of step size, and the points generated by AWS algorithm need to be evenly distributed. In this paper, the step size is uniformly set as 0.01, which is easy to be calculated. If different step sizes are set due to the functional relationship of different optimization objectives, the algorithm is also able to calculate the corresponding results with the minimum step size as the benchmark.

A standard operating condition is set and with lead-acid batteries as the renewable energy transmission equipment for the building. This is achieved without considering the influence of the external environment and other factors. In Case 1, the lack of light is considered; In Case 2, the insufficient wind situation is considered; Case 3 considered the combined effect of wind power and photovoltaic power, electrical connection, and load uncertainty. To show that the INBI algorithm takes into account the tendency of the decision-makers, this paper adopts the weight allocation method in literature [23]. In the new example, the weights of equipment cost, power supply cost, and occupants' comfort are taken as the initial conditions. In Case 4, weight is added based on Case 1, and the weight coefficients are [0.4, 0.3, 0.3]. In Case 5, weight is added based on Case 2, and the weight coefficients are [0.3, 0.4, 0.3]. In Case 6, weight is added based on Case 3, and the weight coefficients are [0.3, 0.3, 0.4]. In Case 7, weight is added based on Case 1, and the weight coefficients are [0.5, 0.25, 0.25]. In Case 8, weight is added based on Case 2, and the weight coefficients are [0.25, 0.5, 0.25]. In Case 9, weight is added based on Case 3, and the weight coefficients are [0.25, 0.25, 0.5]. In Case 10, the lack of light and insufficient wind situation are considered. And the weight coefficients are [1, 1, 1]. In Case 11, the lack of light and insufficient wind situation are considered. And the weight coefficients are [0.4, 0.3, 0.3]. In Case 12, the lack of light and insufficient wind situation is considered, and the weight coefficients are [0.5, 0.25, 0.25]. The details are shown in Table 2.

**Table 2.** Considerations and weight distribution

|  | Considerations | Weight distribution (Equipment cost, Power supply cost, Occupants' comfort) |
|---|---|---|
| Case 1 | the lack of light | [1,1,1] |
| Case 2 | the insufficient wind situation | [1,1,1] |
| Case 3 | the combined effect of wind power and photovoltaic power | [1,1,1] |
| Case 4 | the lack of light | [0.4,0.3,0.3] |
| Case 5 | the insufficient wind situation | [0.3,0.4,0.3] |
| Case 6 | the combined effect of wind power and photovoltaic power | [0.3,0.3,0.4] |
| Case 7 | the lack of light | [0.5,0.25,0.25] |
| Case 8 | the insufficient wind situation | [0.25,0.5,0.25] |
| Case 9 | the combined effect of wind power and photovoltaic power | [0.25,0.25,0.5] |
| Case 10 | the lack of light and insufficient wind situation | [1,1,1] |
| Case 11 | the lack of light and insufficient wind situation | [0.4,0.3,0.3] |
| Case 12 | the lack of light and insufficient wind situation | [0.5,0.25,0.25] |

The INBI algorithm is used in the proposed algorithm. To illustrate that the algorithm proposed in this paper is more accurate, two other algorithms are used in addition to the ones in this paper and are compared. Algorithm 1 is used in the AUAM and the Mahalanobis distance double base point method, and algorithm 2 is used in the original NBI algorithm.

The initial inputs used in the calculation are as follows: The temperature and the light intensity are as shown in Figure 7. The power at the supply side, i.e., the load, is also required, as shown in Figure 8. The power supply includes the power grid, power storage system, photovoltaic power, and wind power. The photovoltaic power and wind power are shown in Figure 9. At the same time, to prove the effectiveness of the proposed method, relevant comparisons are made in the samples. This method can be widely applied to any region of the world and we simply implement this case in China to demonstrate its effectiveness.

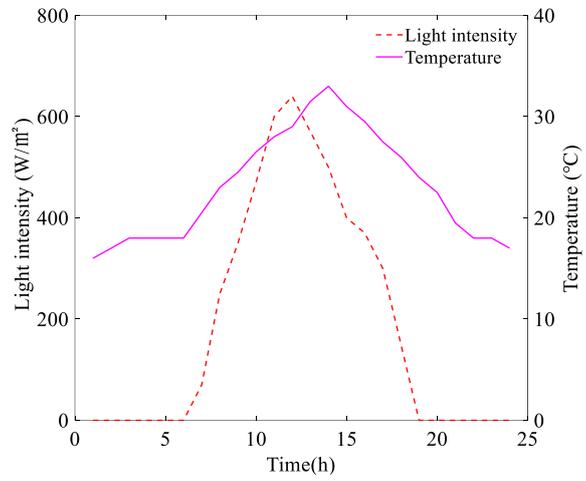

**Fig. 7.** Daily average temperature and light intensity in August

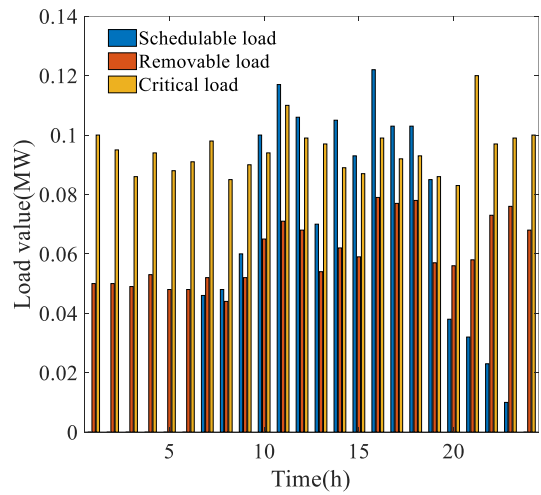

**Fig. 8.** Monthly average load in August

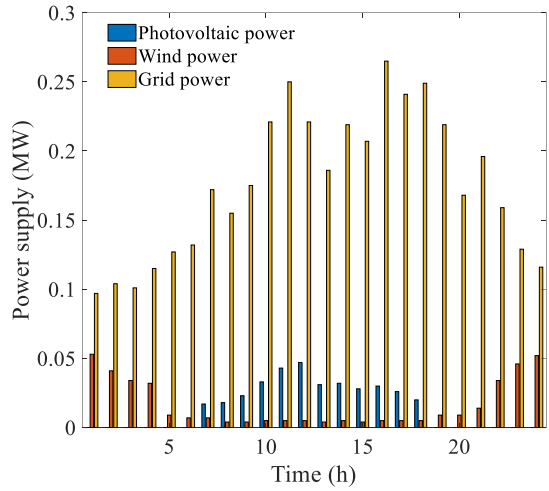

**Fig. 9.** Monthly average power supply in August

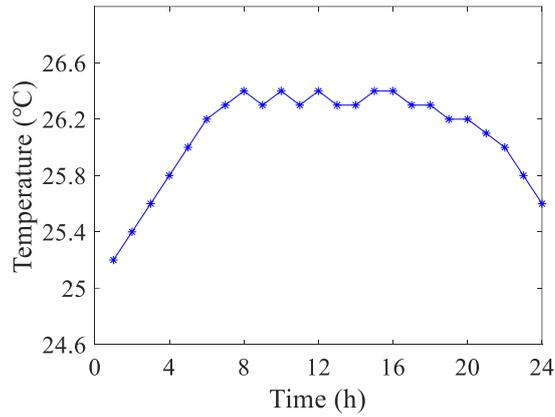

**Fig. 10.** Air temperature changes in the air-conditioned rooms in buildings

Figure 10 shows the indoor temperature changes after turning-ON the air conditioner in the smart building within 24 hours, from which occupants' comfort can be calculated.

*4.2 Calculation result and analysis of smart buildings*

**Table 3.** Calculation results without considering weight distribution

|  | Proposed algorithm | | | | Algorithm 1 | | | | Algorithm 2 | | | |
| --- | --- | --- | --- | --- | --- | --- | --- | --- | --- | --- | --- | --- |
|  | Standard case | Case 1 | Case 2 | Case 3 | Standard case | Case 1 | Case 2 | Case 3 | Standard case | Case 1 | Case 2 | Case 3 |
| Equipment cost (¥) | 3272 | 3377 | 3418 | 3263 | 3587 | 3609 | 3746 | 3591 | 3942 | 4053 | 4099 | 3950 |
| Power | 2225 | 2350 | 2324 | 2202 | 2408 | 2546 | 2518 | 2380 | 2617 | 2755 | 2731 | 2596 |

supply

cost

(¥)

| Occupants' comfort | 94.4% | 93.7% | 93.3% | 93.6% | 92.9% | 92.4% | 92.5% | 92.1% | 91.1% | 90.7% | 90.5% | 90.3% |

**Table 4.** Calculation results considering weight distribution (1)

| | Proposed algorithm | | | | Algorithm 1 | | | | Algorithm 2 | | | |
|---|---|---|---|---|---|---|---|---|---|---|---|---|
| | Standard case | Case 4 | Case 5 | Case 6 | Standard case | Case 4 | Case 5 | Case 6 | Standard case | Case 4 | Case 5 | Case 6 |
| Equipment cost (¥) | 3272 | 3350 | 3439 | 3286 | 3587 | 3583 | 3776 | 3613 | 3942 | 4021 | 4128 | 3982 |
| Power supply cost (¥) | 2225 | 2362 | 2308 | 2215 | 2408 | 2558 | 2503 | 2397 | 2617 | 2772 | 2712 | 2611 |
| Occupants' comfort | 94.4% | 93.7% | 93.2% | 93.7% | 92.9% | 92.4% | 92.5% | 92.2% | 91.1% | 90.6% | 90.5% | 90.4% |

**Table 5.** Calculation results considering weight distribution (2)

| | Proposed algorithm | | | | Algorithm 1 | | | | Algorithm 2 | | | |
|---|---|---|---|---|---|---|---|---|---|---|---|---|
| | Standard case | Case 7 | Case 8 | Case 9 | Standard case | Case 7 | Case 8 | Case 9 | Standard case | Case 7 | Case 8 | Case 9 |
| Equipment cost (¥) | 3272 | 3328 | 3457 | 3303 | 3587 | 3567 | 3793 | 3641 | 3942 | 3995 | 4153 | 4005 |
| Power supply cost (¥) | 2225 | 2389 | 2283 | 2240 | 2408 | 2576 | 2479 | 2415 | 2617 | 2789 | 2698 | 2638 |
| Occupant | 94.4% | 93.6 | 93.1 | 93.8 | 92.9% | 92.3 | 92.4 | 92.3 | 91.1% | 90.5 | 90.4 | 90.5 |

| | s' comfort | % | % | % | | % | % | % | | % | % | % |

Table 6. Calculation results considering weight distribution and weather factors

| | Proposed algorithm | | | | Algorithm 1 | | | | Algorithm 2 | | | |
| --- | --- | --- | --- | --- | --- | --- | --- | --- | --- | --- | --- | --- |
| | Standard case | Case 10 | Case 11 | Case 12 | Standard case | Case 10 | Case 11 | Case 12 | Standard case | Case 10 | Case 11 | Case 12 |
| Equipment cost (¥) | 3272 | 3598 | 3439 | 3396 | 3587 | 3849 | 3641 | 3528 | 3942 | 4205 | 3978 | 3597 |
| Power supply cost (¥) | 2225 | 2475 | 2412 | 2441 | 2408 | 2687 | 2533 | 2593 | 2617 | 2831 | 2799 | 2819 |
| Occupants' comfort | 94.4% | 93.6% | 93.6% | 93.6% | 92.9% | 92.4% | 92.3% | 92.3% | 91.1% | 90.6% | 90.6% | 90.5% |

Tables 3, 4, 5, and 6 show the optimization results of the different algorithms, different weight coefficients, and different cases in the three optimization goals. Compared with Algorithm 1 and Algorithm 2, the number of the Pareto frontier points is significantly improved in this paper. More careful screening of the non-inferior solutions and the compromise solution obtained by this algorithm can reflect the purpose of the decision-makers more effectively. Also, the optimization results are more ideal. Through comparison, it can be seen that the optimization results of the method in this paper are better than those of Algorithm 1 and Algorithm 2.

Table 7. Data comparison between three algorithms in the smart buildings

| | Proposed algorithm | Algorithm 1 | Algorithm 2 |
| --- | --- | --- | --- |
| Equipment cost (¥) | 3349 | 3,651 | 4,033 |
| Power supply cost (¥) | 2,291 | 2,478 | 2,692 |
| Occupants' comfort | 93.6% | 92.4% | 90.6% |

Table 7 shows the average of the results obtained from each algorithm in Tables 3, 4, 5, and 6 under the various operating conditions. It can be seen that, in terms of equipment cost, the proposed algorithm is improved by 8.2% compared with Algorithm 1, and it is increased by 16.9% compared with Algorithm 2. In terms of power supply cost, the proposed algorithm is 7.6% less than in Algorithm 1, and 14.9% better higher than in Algorithm 2. In terms of occupants' comfort, the proposed algorithm is improved by 1.6% compared with Algorithm 1, and it is increased by 3.6% compared with Algorithm 2. Therefore, based on the buildings' optimization, the proposed algorithm is superior to Algorithms 1 and 2 in terms of equipment cost, power supply cost, and occupants' comfort. Taking into account the effects of the combined action of wind power and photovoltaic power, electrical connection, and load uncertainty in smart buildings, the results are optimized. The INBI algorithm uses two Pareto frontier corrections and one accurate selection of the compromise solution to fully consider the correlation between the optimization goals. Moreover, the INBI algorithm takes into account the preference of the decision-makers and improves the applicability of the algorithm.

The proposed method in this paper is an improvement based on the traditional method. It improves its calculation accuracy in solving multi-objective optimization problems. INBI algorithm improves NBI algorithm from three aspects. Firstly, the adaptive weighted sum method (AWS) is used to increase the feasible solution domain and obtain multiple feasible solutions; Secondly, the adjust uniform axis method (AUAM) is used to replace the NBI screening method. 1) The Pareto surface is modified; 2) The more uniform and wide Pareto front are get; The optimal screening step is realized, and the singular points are deleted to the greatest extent. The non-inferior solution is added. As the result, the problem of missing solution is solved; Finally, Mahalanobis distance is used to replace Euclidean distance, and personal preference is considered to add the evaluation matrix, so that the compromise solution is more accurate.

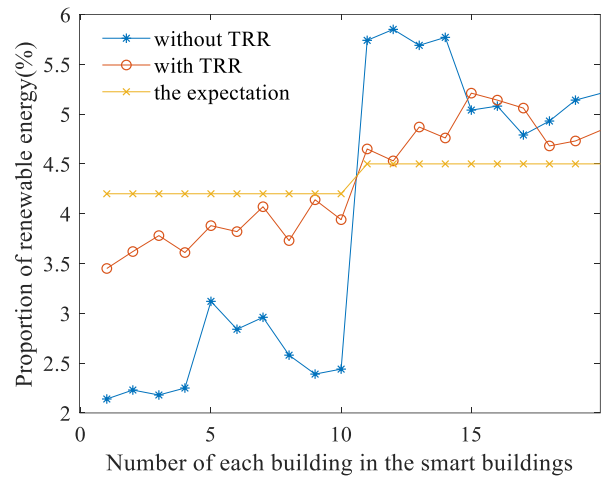

**Fig. 11.** The impact of TRR coefficient on renewable energy allocation

As shown in Figure 11, the building layout adopted in this paper is a pair equation, and the energy consumption law of each building is relatively similar. Decision-makers hope that the energy supply law of each smart building can still be followed after considering the uncertainty of renewable energy and the role of electrical connection. This will make the energy supply of non-renewable energy to smart buildings more stable. It would also reduce the cost of uncertain factors of power supply facilities such as thermal power plants and achieve the purpose of energy optimization in smart buildings. Amongst smart buildings numbered from 1 to 20, 1 to 10 are ordinary buildings, and 11 to 20 are buildings equipped with renewable energy. In smart buildings, the average deviation in renewable energy supply in ordinary buildings and buildings equipped with renewable energy is 0.5% when TRR is integrated into the optimization calculation. This is 1.26% without TRR. In contrast, the average deviation is reduced by 60%. In smart buildings, there is little difference in the actual power consumption of different users. The parameter, TRR, regulates the distributed renewable energy to make the regional renewable energy distribution more uniform and promote the consumption of renewable energy.

*4.3 Characteristic analysis of the smoothing effect in the building control*

The number of buildings involved in the calculation is an important indicator when optimizing the calculation of equipment costs, power supply costs, and occupants' comfort in smart buildings. The number of buildings has an impact on the optimization results. and The optimization results of the individual buildings differ from those of group buildings.

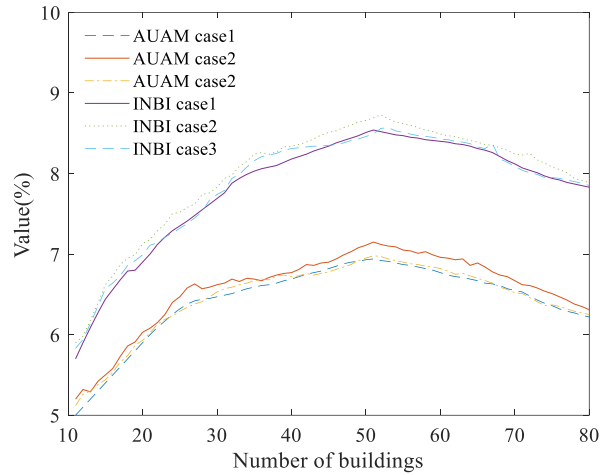

**Fig. 12.** The smoothing effect of smart buildings

With the equipment cost as the calculation target, Figure 12 shows the result of the optimization degree between the proposed algorithm and algorithm 1 in cases 1-3, respectively. The change in optimization is calculated as the number of buildings increases. To explain the problem better, 10 to 80 buildings are selected as a reference in the paper. The two curves in the figure continued to rise until they reached their maximum at 40 buildings and then began to fall.

This phenomenon occurs because the internal adjustment capacity of the building continues to increase with the continuous increase in the number of buildings. Also, the external control measures would continue to reduce the internal energy flow adjustment, which is the smoothing effect of the building. The group buildings' curve is smoother than the individual building's curve. This is because the internal adjustment ability of the group buildings is stronger than that of the individual buildings, and the internal energy is more effectively utilized.

## 6. Conclusions

In this paper, the buildings' controllability and the energy optimization calculation, which includes wind, solar, and air conditioning system, are presented. Taking the equipment cost, power supply cost and occupants' comfort as the optimization target, the results are calculated under the corresponding cases. The conclusions drawn are as follows:

(a) For the first time, the building adjustment assessment index is proposed, named TRR, with wind power, solar power, and air conditioning system. The TRR considers the combined effect of wind power and photovoltaic power. It makes the TRR better interpretive for the distributed energy systems. Furthermore, the TRR constraint and INBI algorithm make the renewable energy allocation in smart buildings more consistent with the intentions of decision-makers.

In smart buildings, the average deviation in renewable energy supply for ordinary buildings and buildings equipped with renewable energy is 0.5% when TRR is integrated into the optimization calculation. This is 1.26% without TRR. In contrast, the average deviation is reduced by 60%.

(b) The NBI algorithm is improved, which is more accurate than the original algorithm optimization. The INBI algorithm uses two Pareto frontier corrections and one accurate selection of the compromise solution to more fully consider the correlation between the optimization goals. Moreover, the INBI algorithm takes into account the preference of decision-makers and improves the applicability of the algorithm.

Compared with the original algorithm, the final result shows that in the three aspects namely: equipment cost, power supply cost, and occupants' comfort, the proposed algorithm is optimized by 8.2%, 7.6%, and 1.6%, respectively.

Acknowledgments:

This work is supported in part by the scientific research project of education department of Liaoning province (LJKZ0129) and science and technology project of state grid corporation of China (SGTYHT/21-JS-226).